\documentclass[a4paper,12pt]{article}

\usepackage {amssymb,amsmath,bbm,graphicx}
\def\b{\begin{eqnarray}}
\def\e{\end{eqnarray}}
\def\bn{\begin{eqnarray*}}
\def\en{\end{eqnarray*}}
\def\>{\rangle}
\def\<{\langle}
\begin{document}
\date{}
\title{Highly Irregular Quantum Constraints}
\author{John R. Klauder\footnote{Also Department of Mathematics}~\footnote{klauder@phys.ufl.edu} and J. Scott Little\footnote{little@phys.ufl.edu}
\\Department of Physics
\\
 University of
Florida\\
 P.O. Box 118440\\ 
Gainesville, FL 32611-8440}
\date{}
\maketitle
\begin{abstract}
Motivated by a recent paper of Louko and Molgado, we consider a simple system with a single classical constraint $R(q)=0$.  If $q_l$ denotes a generic solution to $R(q)=0$, our examples include cases where $R'(q_l)\ne 0$ (regular constraint) and $R'(q_l)=0$ (irregular constraint) of varying order as well as the case where $R(q)=0$ for an interval, such as $a\leq q \leq b$.  Quantization of irregular constraints is normally not considered; however, using the projection operator formalism we provide a satisfactory quantization which reduces to the constrained classical system when $\hbar \rightarrow 0$.  It is noteworthy that irregular constraints change the observable aspects of a theory as compared to strictly regular constraints.
\end{abstract}

\section{Introduction}

The problem discussed in the present paper has been motivated by a recent analysis of the Ashtekar-Horowitz-Boulware (AHB) model \cite{bo} given by Louko and Molgado \cite{lou}.  A brief summary of the AHB model may be given as follows:  $(1.)$ The classical configuration space of the model is taken as a two-dimensional torus with coordinates $(x,y)$  restricted (say) so that $0\leq x,y\leq1$; $(2.)$ With the Hamiltonian for the system set to zero, dynamics is studied in the presence of a constraint given by $p_x^2-R(y)=0$, where $p_x$ is the conjugate momentum to $x$, and $R(y)$ is a suitable function of the second coordinate; $(3.)$  For $p_x$ fixed, the constraint typically restricts the configuration variable $y$ to isolated points, e.g. $y=y_1$, $y=y_2$, etc., a restriction that defines the reduced classical phase space; $(4.)$ Quantization is simplified because the spectrum of $P_x$ (the quantum operator associated with $p_x$) has a discrete spectrum given in our case by $2\pi n$, $n\in\mathbbm{Z}$;  $(5.)$  Attention is focused on imposing the quantum constraint $R(Y) =(2\pi n)^2$, and the difficulties in doing so are qualitatively the same for any $n$;  $(6.)$  For $n=0$ and $R=R_1(y)\equiv(y-c)(y-d)$, $0<c<d<1$, for example, no difficulties arise.  However, for $R=R_2(y)\equiv(y-c)^3(y-d)$, the quantum theory as presented in \cite{lou} involves a physical Hilbert space partitioned into superselection sectors and a classical limit of the reduced quantum theory that does not coincide with the reduced classical phase space referred to above.  Qualitatively speaking, this is the present state of the AHB model. 

In the present paper we examine an even simpler model that carries the same kind of issues as $R_2$ above.  However, we quantize our system by the projection operator method, which is qualitatively different from the refined algebraic quantization method used in \cite{lou}.    Motivated by obtaining a physical Hilbert space that enjoys a classical limit (as $\hbar \rightarrow 0$) which agrees with the original classical reduced phase space, we are able to adapt our procedures not only so as to achieve this goal but at the same time find that no imposition of superselection sectors is required. As a matter of notation, we generally do not make a distinction between the constraint hypersurface and that surface modulo any gauge action; the context will make clear which space is under consideration.

The model we choose to study involves just one configuration variable $q$, $-\infty<q<\infty$, and its conjugate variable $p$.  The classical action is taken to be
\begin{equation}
I=\int [p\dot{q}-\lambda R(q)]dt,
\end{equation}
where $\lambda$ is a Lagrange multiplier designed to enforce the single constraint
\begin{equation}
R(q)=0\;.
\end{equation}
(The analog to the AHB model is that this model examines just the ``$n=0$ sector''.)  The classical equations of motion for our simple system are given by
\b
\dot{q}&=&0,\\
\dot{p}&=&-\lambda R'(q),\\
R(q)&=&0\;,
\e
with solutions
\b
q(t) &=& q_l = q(0), \\
p(t) &=& -R'(q_l)\int_0^t \lambda(t')dt' +p(0),
\e
where $q_l$ is a root of $R(q)=0$. (Restrictions on the class of functions $R$ are given in the next section.)  If $R'(q_l)=0$ then the solution becomes

\begin{equation}
q(t)=q_l=q(0),\hskip.5cm p(t)=p(0).
\end{equation}

The reduced classical phase space is given by $\mathbbm{R} \times {\cal Z}$, where
\begin{equation}
{\cal Z} =\{q:  R(q)=0\}.
\end{equation}
The function $\lambda (t)$ is not fixed by the equations of motion, which is normal for systems with first-class constraints. To explicitly exhibit a solution to the classical equations of motion it is generally necessary to specify the function $\lambda (t)$, and this constitutes a choice of gauge.  Gauge dependent quantities are defined to be unobservable, while gauge independent quantities are declared to be observable.  In the present example, if $R'(q_l)\ne 0$, then $p(t)$ is gauge dependent, while if $R'(q_l)=0$, $p(t)$ is, in fact, gauge independent.  This behavior suggests that the momentum $p$ in the subset of the reduced classical phase space for which $\{q: R(q)=0, \, R'(q)\ne 0\}$ is unobservable, while the momentum $p$ in the subset of the reduced classical phase space for which $\{q: R(q)=0, \, R'(q)= 0\}$ is observable. We discuss this point further below.

Such a situation is generally considered undesirable, and most treatments of problems involving constraints require that the constraints be ``regular'' \cite{gau}, which for us means that $R'(q_l)\ne0$ always so that the variable $p$ would always be gauge dependent.  The example $R_1(q)=(q-c)(q-d)$, now with $-\infty<c<d<\infty$, leads to regular constraints at $q=c$ and $q=d$.  When $R'(q_l)=0$, on the other hand, the constraint is called ``irregular'', and it typically requires special treatment, if it can be treated at all.  The example $R_2(q)=(q-c)^3(q-d)$ leads to a regular constraint at $q=d$ and an irregular constraint at $q=c$ (with an ``order 3'' power).  Strange as it may seem, the two examples based on $R_1(q)$ and $R_2(q)$ lead to the same classical reduced phase space (such as it is) composed of the two lines $(p,c)$ and $(p,d)$ -- but they entail different observability properties: For $R_1$, $p\in(p,c)$ and $p\in(p,d)$ are both unobservable, while for $R_2$, $p\in (p,c)$ is observable and $p\in (p,d)$ is unobservable.

We are led to the conclusion that the set of observables is not determined by just the reduced classical phase space.  Of course, one can always choose a dynamical gauge and thereby uniquely fix $p$, e.g., $p=p^G$, a constant, after which $p$ becomes observable (by being uniquely determined), and this is a common procedure to eliminate such problems.  However, fixing a gauge is to be resisted if at all possible since doing so then requires a proof that the final results obtained do not depend on the particular gauge choice made, and such a proof is often difficult or even impossible to establish.

Constraint functions involving both regular and irregular, or two or more irregular constraints of differing order, may be called ``highly irregular''.  While such constraint expressions may be unwelcome, they nevertheless may well arise and deserve to be considered.  Our goal in this paper is to discuss some of the issues related to the introduction and analysis of highly irregular quantum constraints by means of the projection operator formalism \cite{klauder}.

The paper is organized as follows:  Section 2 presents a quantum treatment of a generalized version of our model resulting in a characterization of the resultant physical Hilbert space. Section 3 provides a brief introduction to observables in the projection operator formalism and establishes that the physical Hilbert space is not encumbered with any superselection requirements.  Section 4 deals with the classical limit of the constrained physical quantum theory and establishes that the classical limit includes the complete reduced classical phase space of the original model.  

In a subsequent paper, we shall apply the methods and results of the present paper to study the original AHB model itself.  This study will show that this model can be quantized in such a way that the physical Hilbert space is not divided into superselection sectors and the classical limit of the reduced quantum theory agrees with the original reduced classical theory.

\section{Quantization} 
Let us first describe the model of principal interest.  To that end we initially remark that
if $q=q_l$ is a solution of $R(q)=0$, then it follows that 
\b
q(t) &=& q_l = q(0), \\
p(t) &=& -R'(q_l)\int_0^t \lambda(t')dt +p(0),
\e
where $q(0)$ and $p(0)$ are the initial values.  In particular, if $R'(q_l)=0$ then
\begin{equation}
q(t)=q_l=q(0), \hskip .5cm p(t)=p(0).
\end{equation} 
Clearly, for the classical theory to be well defined, it is sufficient for $R(q) \in C^1$, namely that $R(q)$ and $R'(q)$ are both continuous.  (Strictly speaking this continuity is required only in the neighborhood of the zero set $\{q: R(q)=0\}$; however, with an eye toward the AHB model we choose $R(q)\in C^1$ for all $q$.)

Our discussion will cover a wide class of $R$ functions, and for convenience of explanation we shall focus on one specific example; generalization to other examples is immediate.  The example we have in mind is given by 
\begin{equation}
R(q)=q^3(1-q)(q-2)^2(3-q)^{3/2}\theta(3-q) + e^{-\frac{1}{(q-4)}}\theta(q-4),
\end{equation} 
where
\begin{equation}
\theta(x)\equiv
\begin{cases}
1,& x>0\\
0,& x<0.
\end{cases}
\end{equation}
For this example, the zero set is given by 
\begin{equation}
{\cal Z}\equiv \{q=0,\, q=1,\, q=2,\, 3\leq q \leq 4\};
\end{equation}
only for $q=1$ is $R'(q)\ne 0$.
(Although physically motivated models would typically not include intervals 
in the zero set of $R$, we do so to illustrate the versatility of our 
approach.)

In summary, the phase space for the unconstrained classical system is parameterized by $(p,q) \in \mathbbm{R}\times \mathbbm{R}$, and the phase space for the constrained system is parameterized by the points $(p,q)\in  \mathbbm{R}\times {\cal Z}$.  This latter space consists of several one-dimensional lines and a two-dimensional strip.  From the standpoint of this elementary example all elements of $\mathbbm{R}\times {\cal Z}$ are equally significant.

We now turn to the quantization of this elementary example following the precepts of the projection operator formalism \cite{klauder}.  In this approach one quantizes first and reduces second.  The ultimate reduction leads to a physical Hilbert space appropriate to the constrained system.  As a check on our procedures we require that the classical limit ($\hbar\rightarrow 0$) of the quantized constrained system recovers the classical constrained system with phase space $\mathbbm{R}\times {\cal Z}$ as described above.

Quantization first means that our original variables $p$ and $q$ become conventional self-adjoint operators $P$ and $Q$ subject to the condition that 
\begin{equation}
[Q,P]= i \mathbbm{1}
\end{equation}
in units where $\hbar=1$.  (When we eventually examine the classical limit, we shall restore the parameter $\hbar$ to various expressions as needed.)  The projection operator of interest is given by 
\begin{equation}
\mathbbm{E}(R(Q)^2\leq\delta^2)=\mathbbm{E}(-\delta\leq R(Q)\leq\delta)=\mathbbm{E}(-\delta< R(Q)<\delta)\;,
\end{equation}
where $\delta>0$ is a temporary regularization parameter that will eventually be sent to zero in a suitable manner.  Since the limit $\delta\rightarrow0$ will ultimately be taken as a form limit, we need to introduce suitable bras and kets in this original, unconstrained Hilbert space.  For that purpose we will choose canonical coherent states defined, for the present discussion, by
\begin{equation}
|p,q\>\equiv e^{ipQ}e^{-iqP}|0\>.
\end{equation}
As usual, we choose $|0\>$ to satisfy $(Q+iP)|0\> =0$; namely, $|0\>$ is the normalized ground state of an harmonic oscillator with unit frequency and mass.  Thus we are led to consider the complex function
\begin{equation}
\<p'',q''|\mathbbm{E}(-\delta<R(Q)<\delta)|p',q'\> \label{pt}
\end{equation}
which is continuous (actually $C^\infty$) in the coherent state labels and uniformly bounded by unity since $\mathbbm{E}=\mathbbm{E^\dagger}=\mathbbm{E}^2 \leq \mathbbm{1}$.

It is important to remark that the function (\ref{pt}) is a function of positive type, a criterion that means  
\begin{equation}
\Sigma^N_{j,k=1} \alpha^*_j\alpha_k \<p_j,q_j|\mathbbm{E}|p_k,q_k\> \geq 0
\end{equation}
for all $N<\infty$ and arbitrary complex numbers $\{\alpha_j\}$ and label sets $\{p_j,q_j\}$;  this property holds because $\mathbbm{E}$ is a projection operator.  As a consequence of being a continuous function of positive type the function 
\begin{equation}
{\cal K}(p'',q'';p',q')\equiv\<p'',q''|\mathbbm{E}|p',q'\> \label {rk}
\end{equation} 
serves as a reproducing kernel for a reproducing kernel Hilbert space ${\cal H}$, a functional representation by continuous functions on the original phase space ($\mathbbm{R} \times \mathbbm{R}$), of the regularized (by $\delta >0$) physical Hilbert space.  Our goal is to take a suitable limit $\delta \rightarrow 0$ so as to yield a function that can serve as a reproducing kernel for the true physical Hilbert space for the present problem.

Clearly the limit $\delta \rightarrow 0$ of the given expression vanishes and that is an unacceptable result.  Suppose we assume $0< \delta \ll 1$, e.g., $\delta = 10^{-1000}$.  Then it is clear (even for a much larger $\delta$ as well!), for the example at hand, that 
\b
\nonumber \mathbbm{E}(-\delta< R(Q)<\delta) &= &  \mathbbm{E}(-\delta<12\sqrt{3}Q^3<\delta)\\
\nonumber &+&\mathbbm{E}(-\delta< 2\sqrt{2}(1-Q)<\delta)+\mathbbm{E}(-\delta<-8(Q-2)^2 <\delta)\\
\nonumber &+& \mathbbm{E}(-\delta<-54(3-Q)^{3/2}<0; Q<3) + \mathbbm{E}(3\leq Q\leq 4)\\
\nonumber &+&\mathbbm{E}(0<e^{-\frac{1}{(Q-4)}} <\delta; Q>4) \\
&\equiv& \mathbbm{E}_1+\mathbbm{E}_2+\mathbbm{E}_3+\mathbbm{E}_4+\mathbbm{E}_5+\mathbbm{E}_6\;,
\e
where $\mathbbm{E}_n, 1\leq n\leq6$, corresponds to the terms in the line above in order.  By construction, for very small $\delta$, it follows that these projection operators obey
\begin{equation}
\mathbbm{E}_n\mathbbm{E}_m =\delta_{nm}\mathbbm{E}_n,
\end{equation}
i.e., they project onto mutually orthogonal subspaces.  In like manner the reproducing kernel decomposes into 
\b
\nonumber {\cal K}(p'',q'';p',q') &=& \<p'',q''|\mathbbm{E}(-\delta< R(Q)<\delta)|p',q'\>\\
&=& \Sigma^6_{n=1}{\cal K}_n(p'',q'';p',q'),
\e
where
\begin{equation}
{\cal K}_n(p'',q'';p',q') \equiv \<p'',q''|\mathbbm{E}_n|p',q'\>.
\end{equation}
Each function ${\cal K}_n(p'',q'';p',q')$ serves as a reproducing kernel for a reproducing kernel Hilbert space ${\cal H}_n$, and the full reproducing kernel Hilbert space is given by  
\begin{equation}
{\cal H}= \bigoplus^6_{n=1} {\cal H}_n.
\end{equation} 
 Since $\mathbbm{E}_n\mathbbm{E}_m =\delta_{nm}\mathbbm{E}_n$ it follows, from the completeness of the coherent states, that 
\begin{equation}
\int {\cal K}_n(p'',q'';p,q){\cal K}_m(p,q;p',q')dpdq/(2\pi) = \delta_{nm}{\cal K}_n(p'',q'';p',q'). \label{t27}
\end{equation}
This equation implies that the ${\cal H}_n$, $1\leq n \leq 6$, form 6 mutually disjoint (sub) Hilbert spaces within $L^2(\mathbbm{R}^2)$. For the present example with $\delta>0$, each ${\cal H}_n$ is infinite dimensional.
  
Let us first consider 
\bn
 {\cal K}_2(p'',q'';p',q')&\equiv&\<p'',q''|\mathbbm{E}(-\delta< 2\sqrt{2}(1-Q)<\delta)|p',q'\>\\
&=&\frac{1}{\sqrt{\pi}}\int^{1+\delta_2}_{1-\delta_2} e^{-(x-q'')^2/2 -i(p''-p')x}e^{-(x-q')^2/2} dx\;,
\en
where $\delta_2 \equiv \delta/(2\sqrt{2})$.  This expression manifestly leads to a function of positive type.  For very small $\delta$ (e.g., $\delta = 10^{-1000}$
), we can assert that to leading order
\begin{equation} 
{\cal K}_2(p'',q'';p',q')= \frac{2\delta_2}{\sqrt{\pi}} e^{-(1-q'')^2/2 -i(p''-p')-(1-q')^2/2}\frac{\sin \delta_2(p''-p')}{\delta_2(p''-p')}.
\end{equation}
This function is already of positive type, exactly fulfills (\ref{t27}), and is correct to ${\sf O}(\delta^2)$ [i.e., to ${\sf O}(10^{-2000})$!].

As discussed frequently before \cite{klauder}, we can extract the ``germ" from this reproducing kernel by first scaling it by a factor of ${\sf O}(\delta^{-1})$, say by $\sqrt{\pi}/(2\delta_2)$, prior to taking the limit $\delta\rightarrow 0$.
Consequently, we first define a new reproducing kernel
 \begin{equation}
{\hat {\cal K}}_2(p'',q'';p',q')=\frac{\sqrt{\pi}}{2\delta_2} {\cal K}_2(p'',q'';p',q').
\end{equation}
We remark that the space of functions that make up the reproducing kernel Hilbert space  ${\hat{\cal H}}_2$ (generated by ${\hat {\cal K}}_2$) is {\it identical} to the space of functions that make up the reproducing kernel Hilbert space ${\cal H}_2$ (generated by ${\cal K}_2$); of course, there is a difference in the inner product assigned to the two Hilbert spaces ${\hat{\cal H}}_2$ and ${\cal H}_2$, each of which is given by the fact that for each element in a generic reproducing kernel Hilbert space given by
\b     \psi(p,q)\equiv\Sigma_{j=1}^N \alpha_j\,K(p,q;p_j,q_j)\;,  \e
the norm square of that element is given by
 \b \|\psi\|^2=(\psi,\psi)\equiv\Sigma_{j,k=1}^N\alpha^*_k\alpha_j\,K(p_k,q_k;p_j,q_j)\;, \label{e4} \e
where of course $K(p'',q'';p',q')$ denotes the generic 
reproducing kernel under consideration.

Next, we take the limit as $\delta\rightarrow 0$ of the function ${\hat {\cal K}_2}$, which leads to 
\b
\nonumber {\tilde {\cal K}}_2(p'',q'';p',q') &\equiv& \lim_{\delta \to 0}{\hat {\cal K}}_2(p'',q'';p',q')\\
&=& e^{-[(1-q'')^2+(1-q')^2]/2 - i(p''-p')}. 
\e
This procedure leads to a new function ${\tilde {\cal K}_2}$, which, provided it is still continuous -- which it is -- leads to what is called a reduced reproducing kernel and thereby also to a new reproducing kernel Hilbert space ${\tilde {\cal H}}_2$ .  Generally, the dimensionality of the space as well as the definition of the inner product are different for the new reproducing kernel Hilbert space; however, one {\it always} has the standard inner product definition that is appropriate for any reproducing kernel Hilbert space \cite{RKHS}, as follows from (\ref{e4}).  In the present case, it follows that ${\tilde {\cal K}}_2$ defines a {\it one}-dimensional Hilbert space ${\tilde {\cal H}}_2$  .  Note that even though the coordinate value for the constrained coordinate $Q$ is now set at $Q=1$ --~as is clear from the special dependence of ${\tilde {\cal K}}_2(p'',q'';p',q')$ on $p''$ and $p'$, as well as the fact that $-i\partial/\partial p'$ acting on ${\tilde{\cal K}}_2(p'',q'';p',q')$  generates matrix elements of $Q$~-- the range of the values $q''$ and $q'$ is still the whole real line.  The only remnant that $q''$ and $q'$ have of their physical significance is that ${\tilde {\cal K}}_2(p'',q'';p',q')$ peaks at $q''=q'=1$.

A similar procedure is carried out for the remaining components  in the original reproducing kernel.  Let us next consider
\bn
 {\cal K}_1(p'',q'';p',q')&=&\<p'',q''|\mathbbm{E}(-\delta< 12\sqrt{3}Q^3<\delta)|p',q'\>\\
&=&\frac{1}{\sqrt{\pi}}\int^{\delta_1}_{-\delta_1} e^{-(x-q'')^2/2 -i(p''-p')x}e^{-(x-q')^2/2} dx
\en
where $\delta_1 \equiv [\delta/(12\sqrt{3})]^{1/3}$. To leading order 
\begin{equation} 
{\cal K}_1(p'',q'';p',q')= \frac{2\delta_1}{\sqrt{\pi}} e^{-(q''^2+ q'^2)/2}\frac{\sin \delta_1(p''-p')}{\delta_1(p''-p')},
\end{equation}
which is a function of positive type that exactly fulfills (\ref{t27}).  It is noteworthy that an example of this type of irregular constraint was considered previously by one of us (JRK) \cite{klauder2}.

We rescale this function differently so that 
 \begin{equation}
{\hat {\cal K}}_1(p'',q'';p',q')\equiv\frac{\sqrt{\pi}}{2\delta_1} {\cal K}_1(p'',q'';p',q')
\end{equation}
and then take the limit $\delta \rightarrow 0$ leading to 
\b
\nonumber {\tilde {\cal K}}_1(p'',q'';p',q') &\equiv& \lim_{\delta \to 0}{\hat {\cal K}}_1(p'',q'';p',q')\\
&=& e^{-[q''^2+q'^2]/2}, 
\e
a continuous function of positive type that characterizes the one-dimensional Hilbert space ${\tilde {\cal H}}_1$.

Our procedure of scaling the separate parts of the original reproducing kernel by qualitatively different factors (i.e., $\delta_1$ and $\delta_2$) has not appeared previously in the projection operator formalism.  This difference in scaling is motivated by the goal of having each and every element of the reduced classical phase space represented on an equal basis in the quantum theory.  It is only by this procedure that we can hope that the classical limit of expressions associated with the physical Hilbert space can faithfully recover the physics in the classical constrained phase space.  Scaling of ${\tilde {\cal K}_1}$ and ${\tilde{\cal K}_2}$ by finitely different factors is taken up later.

Let us continue to examine the remaining ${\cal K}_n$, $3\leq n\leq 6$.  For ${\cal K}_3$ we have
\bn
 {\cal K}_3(p'',q'';p',q')&=&\<p'',q''|\mathbbm{E}(-\delta< -8(Q-2)^2<\delta)|p',q'\>\\
&=&\frac{1}{\sqrt{\pi}}\int^{2+\delta_3}_{2-\delta_3} e^{-(x-q'')^2/2 -i(p''-p')x}e^{-(x-q')^2/2} dx,
\en
where $\delta_3 \equiv [\delta/8]^{1/2}$.  The now familiar procedure leads to 
\begin{equation}
{\tilde {\cal K}}_3(p'',q'';p',q') =  e^{-[(2-q'')^2+(2-q')^2]/2 - 2i(p''-p')}
\end{equation}
corresponding to a one-dimensional Hilbert space ${\tilde {\cal H}}_3$. For ${\cal K}_4$ we find that 
\bn
 {\cal K}_4(p'',q'';p',q')&=&\<p'',q''|\mathbbm{E}(-\delta< -54(3-Q)^{3/2}<\delta; Q<3)|p',q'\>\\
&=&\frac{1}{\sqrt{\pi}}\int^{3}_{3-\delta_4} e^{-(x-q'')^2/2 -i(p''-p')x}e^{-(x-q')^2/2} dx,
\en
where $\delta_4 \equiv [\delta/54]^{2/3}$; thus we may choose 
\begin{equation}
{\tilde {\cal K}}_4(p'',q'';p',q') =  e^{-[(3-q'')^2+(3-q')^2]/2 - 3i(p''-p)},
\end{equation}
and so ${\tilde {\cal H}}_4$ is one-dimensional.  For ${\cal K}_5$ we are led to \b
\nonumber  {\cal K}_5(p'',q'';p',q')&=&\<p'',q''|\mathbbm{E}(3\leq Q \leq 4)|p',q'\>\\
&=&\frac{1}{\sqrt{\pi}}\int^{4}_3 e^{-(x-q'')^2/2 -i(p''-p')x}e^{-(x-q')^2/2} dx. \label{K5}
\e
In this case, no $\delta$ appears and no infinite rescaling is needed, so we may simply choose
 \begin{equation}
{\tilde {\cal K}}_5(p'',q'';p',q')={\cal K}_5(p'',q'';p',q').
\end{equation}
Although we do not have an explicit analytic expression for ${\tilde {\cal K}_5}$ , we do have a well-defined integral representation in (\ref{K5}). Furthermore, it follows that ${\tilde {\cal H}_5}$ is infinite dimensional.

Lastly, for ${\cal K}_6$ we see that 
\bn
 {\cal K}_6(p'',q'';p',q')&=&\<p'',q''|\mathbbm{E}(0< e^{-\frac{1}{(Q-4)}}<\delta; Q>4)|p',q'\>\\
&=&\frac{1}{\sqrt{\pi}}\int^{4+\delta_6}_{4} e^{-(x-q'')^2/2 -i(p''-p')x}e^{-(x-q')^2/2} dx,
\en
where $\delta_6 \equiv 1/ |\ln(\delta)|$ , and we let 
\begin{equation}
{\tilde {\cal K}}_6(p'',q'';p',q') =  e^{-[(4-q'')^2+(4-q')^2]/2 - 4i(p''-p')}
\end{equation}
corresponding to a one-dimensional space ${\tilde {\cal H}}_6.$

Finally, we define the reproducing kernel for the physical Hilbert space as
\begin{equation}
{\tilde {\cal K}}(p'',q'';p',q')  \equiv \Sigma^6_{n=1}{\tilde {\cal K}}_n(p'',q'';p',q')\;.
\end{equation}
In turn, the physical Hilbert space ${\cal H}_P$ is defined as the reproducing kernel Hilbert space ${\tilde {\cal H}}$ uniquely determined by the reproducing kernel ${\tilde {\cal K}}(p'',q'';p',q')$. 

Observe, by our procedure, all elements of the reduced classical phase space ($\mathbbm{R} \times {\cal Z}$)  are represented on an equivalent basis in ${\tilde {\cal K}}$ .  This feature has been designed so that the classical limit of the expressions within ${{\cal H}_P}$  correspond to all aspects of the reduced classical phase space.  After a discussion of observables for our model problem, we will discuss the classical limit of our physical Hilbert space construction.
\section{Observables}
Let us restrict our discussion of observables to those that are self-adjoint operators ${\cal O}$ in the unconstrained Hilbert space.  We also limit the discussion to constraints that are both classically and quantum mechanically first class. 

We first discuss the situation in the case of a regularized ($\delta>0$) enforcement of the constraints.  In this case, all physical observables must obey the following identity
\begin{equation}
 [\mathbbm{E}, {\cal O}]=0. \label{EO}
\end{equation} 
We can take a general ${\cal G}(P,Q)$  in the unconstrained Hilbert space and define
\begin{equation}
{\cal G}^E(P,Q)\equiv \mathbbm{E}{\cal G}(P,Q)\mathbbm{E}. \label{obs}
\end{equation}
as its observable component since clearly $[\mathbbm{E}, {\cal G}^E] =
0$. In fact, every observable can be expressed in the preceding form. Equation (\ref{EO}) is valid for $\delta>0$; however, as long as $\delta>0$, we have yet to capture the true physical Hilbert space of a given theory.  Therefore the limit $\delta \rightarrow 0$ must be taken in a suitable fashion to discuss observables.   
 
If the set of constraints admits a discrete spectrum that includes zero --~which is {\it not} the case for our particular model~-- then we can make the following claims.  An operator ${\cal O}$ is observable if the following is valid

\begin{equation}
\lim_{\delta \to 0} \, [\mathbbm{E}, {\cal O}]=0 \hskip.25cm \Leftrightarrow \hskip.25cm [\Phi,{\cal O}]|\psi\>_{Phys}=0, \label{iff}
\end{equation}
where $\Phi$ denotes any one of the constraint operators.

The first part of the preceding if and only if statement has no classical analog; however, the second statement is related to the following weak classical equation
\begin{equation}
\{\phi, o\} \approx 0, \label{weak}
\end{equation}
where the symbols $\phi$ and  $o$ denote the classical version of the 
constraint and observable, respectively.
We consider (\ref{weak}) to be a weak equation because it need only vanish on the constraint hypersurface.  It is obvious that if (\ref{iff}) is true then ${\cal O}$ is gauge independent in the physical Hilbert space.  In the Heisenberg picture the evolution of the operator is given by 

\begin{equation}
{\dot{\cal O}}|\psi\>_{Phys}=i\,[H^E, {\cal O}]|\psi\>_{Phys}\;,
\end{equation}  
where $H^E$ is the observable part of the Hamiltonian in the form of  (\ref{obs}).  Therefore all observables will stay in the physical Hilbert space as they evolve with time.  The same type of statement can also be said in the classical world.  However, in our particular model, the limit $\delta\rightarrow 0$ must be taken as a form limit because the constraint has its zero in the continuous spectrum.  Observables in these instances must be handled at the level of the reduced reproducing kernel.  

Before discussing the observables in our particular model, let us first determine the form of the projection operator for the physical Hilbert space.  As shown in the previous section, the physical Hilbert space, for the specifically chosen model, can be written as the direct sum of 5 one-dimensional Hilbert spaces direct summed with an infinite (separable) Hilbert space.  Introducing a complete orthonormal basis for ${\tilde {\cal H}_5}$, it follows that the physical Hilbert space is isomorphic to an infinite direct sum of complex numbers,
\begin{equation}
{{\cal H}_P}= \oplus^{\infty}_{l=1} \mathbbm{C}_l.
\end{equation}
In this realization the projection operator is merely the unit matrix, and thus observables correspond to general Hermitian matrices.

We now direct our attention to a calculation of the  coherent state matrix elements of the physical conjugate momentum at the level of the reproducing kernel. Specifically, we first note that 
\bn
\<p'',q''|P^E|p',q'\> &\equiv& \<p'',q''|\mathbbm{E}P\mathbbm{E}|p',q'\>\\
&=& \int dx \int dx' \<p'',q''|\mathbbm{E}|x\>\<x|P|x'\>\<x'|\mathbbm{E}|p',q'\>\\
&=& -i\hbar\int dx \int dx' \<p'',q''|\mathbbm{E}|x\>\delta'(x-x')\<x'|\mathbbm{E}|p',q'\>\\
&=& -i\hbar \int dx \<p'',q''|\mathbbm{E}|x\>\frac{d}{dx}\<x|\mathbbm{E}|p',q'\>.
\en
We implement the constraint by integrating over the appropriate intervals $\{I_n\}$ implicit in (22),  \begin{equation}
\<p'',q''|P^E|p',q'\>= -i\hbar \Sigma_n\int_{I_n} dx \<p'',q''|x\>[ip'/\hbar+ (q'-x)/\hbar]\<x|p',q'\>. \label{matrix}
\end{equation}
Similarly, it follows that
\begin{equation}
\<p',q'|P^E|p'',q''\>^*= i\hbar \Sigma_n \int_{I_n} dx \<p'',q''|x\>[-ip''/\hbar+ (q''- x)/\hbar]\<x|p',q'\>. \label{matrix2}
\end{equation} 
Following Araki \cite{?}, we now determine the desired matrix elements by adding (\ref{matrix}) and (\ref{matrix2}), and dividing by two, which leads to
\b
\nonumber \<p'',q''|P^E|p',q'\> &=& \frac{1}{2}((\ref{matrix}) +(\ref{matrix2}))\\
\nonumber &=&\frac{1}{2}\Sigma_n\int_{I_n} dx \<p'',q''|x\>[p''+p'+ i(q''- q')]\<x|p',q'\>\\
\nonumber &=& \frac{p''+p'+ i(q''- q')}{2}\Sigma_n\int_{I_n}\<p'',q''|x\>\<x|p',q'\>dx \\
&=&  \frac{p''+p'+ i(q''- q')}{2} {{\cal K}} (p'',q'';p',q'). \label{mo1}
\e

Finally, if we so choose, we may fix the gauge dependent matrix elements by gauge fixing the portions of the reproducing kernel that correspond to the regular constraints (i.e. $R'(q_l)\neq 0$).  For our example, this leads to a new
reduced reproducing kernel given by
\b
   {\cal K}'(p'',q'';p',q')=\Sigma_{n\,(\ne 2)}{\cal K}_n(p'',q'';p',q')+{\cal K}_2(p^G,q'';p^G,q')\;, 
\e
where $p=p^G$ is the constant gauge choice. Moreover, for the physical matrix elements of $P$ we are led to
\b
 \<p'',q''|P^E|p',q'\>& =& \frac{p''+p'+ i(q''- q')}{2} \Sigma_{n\,(\ne 2)}{\cal K}_n(p'',q'';p',q')\nonumber\\
&& + \frac{2p^G+ i(q''- q')}{2}{\cal K}_2(p^G,q'';p^G,q')\label{mo2}\;.
\e
Rescaling and the limit $\delta\rightarrow0$ proceed in direct analogy with
the discussion in Sec. 2.

\section{Classical Limit}

Let us begin our discussion of the classical limit with the original, unconstrained quantum system.  While our previous discussion could have been based on choosing among a great variety of total sets of bras and kets with which to form our matrix elements, there is a clear advantage to having chosen coherent states as our bras and kets.  The advantage in this choice stems from the remarkable felicity with which coherent states may be used to interpolate between the quantum and classical theories.  

In the original, unconstrained Hilbert space the general rule given by the
diagonal coherent state matrix elements, namely
\begin{equation}
\<p,q|{\cal G}(P,Q)|p,q\> = G(p,q;\hbar)
\end{equation}
establishes a connection between a general quantum operator ${\cal G}(P,Q)$ and an associated function (or symbol) on the classical phase space.  The function $ G(p,q;\hbar)$ is generally not the classical function associated with  ${\cal G}(P,Q)$ for the simple reason that $\hbar>0$ .  However, in the limit $\hbar \rightarrow 0$  we find that
\bn
\lim_{\hbar \to 0}\<p,q|{\cal G}(P,Q)|p,q\> &=&\lim_{\hbar \to 0}\<0|{\cal G}(p+P,q+Q)|0\>\\
&\equiv& {\cal G}(p,q),
\en
 which is readily seen to hold when ${\cal G}$ is a polynomial, provided $\lim_{\hbar \to 0} \<0|(P^2 +Q^2)|0\> =0$, which is clearly true for our choice of $|0\>$; of course, this result has assumed that the only dependence on $\hbar$ arises from that of $P$ and $Q$. Besides the elementary example that the unit operator ${\cal G} = \mathbbm{1}$  images to $G=1$, the 
next simplest examples are
\begin{equation}
\<p,q|Q|p,q\> = q, \hskip .5cm   \<p,q|P|p,q\> = p,
\end{equation}
expressions which remain unchanged as $\hbar \rightarrow 0$ .  These relations establish the basic fact that the range of the classical coordinate and momentum coincides with the range of the corresponding labels of the coherent states namely, $(p,q)\in\mathbbm{R}\times\mathbbm{R}$. (This self-evident observation will be relevant when we discuss the constrained system.)

Clearly, the unconstrained classical phase space is invariant under translations, such as $q\rightarrow q+a$ and/or $p\rightarrow p+b$, where $-\infty<a,b<\infty$.  Such translations can be generated by macroscopic canonical transformations.  Let $\{q,p\} = 1$ denote the fundamental Poisson bracket and define
\b
\{q,\cdot\}H(p,q) &\equiv& \{q, H(p,q)\},\\
\{p,\cdot\}H(p,q) &\equiv& \{p, H(p,q)\},
\e
for an analytic phase space function $H(p,q)$. Then it follows that 
\b
S_aH(p,q) &\equiv& e^{-a\{p,\cdot\}}H(p,q)\nonumber\\
&\equiv& \Sigma^\infty_{m=0} \frac{(-a)^m}{m!}\{p,\{p,\cdots \{p, H(p,q)\} \cdots \} \} \nonumber\\
&=& H(p,q+a)\label{qwer}
\e
and
\b
T_aH(p,q) &\equiv& e^{b\{q,\cdot\}}H(p,q)\nonumber\\
&\equiv& \Sigma^\infty_{m=0} \frac{b^m}{m!}\{q,\{q,\cdots \{q, H(p,q)\} \cdots \} \} \nonumber\\
&=& H(p+b,q).
\e
We may also adopt the point of view that $S_a$ and $T_b$ are translation transformations defined simply by asserting that 
\b
S_aH(p,q) &\equiv& H(p,q+a)\;,\\
T_bH(p,q) &\equiv& H(p+b,q)
\e
for general phase space functions $H(p,q)$.  

We next study translation transformations acting on the constrained classical phase space  $\mathbbm{R} \times {\cal Z}$.  Let $F: \mathbbm{R} \times {\cal Z} \rightarrow \mathbbm{R}$ be a function on the constrained classical phase space.  This means that $F(p,q)$ is defined only on  $\mathbbm{R} \times {\cal Z}$ and nowhere else.  Translations in $p$ are unaffected by the constraint so we focus only on translations in $q$.  The transformation $S_a$ is defined for $F(p,q)$ and acts as 
\begin{equation}
S_aF(p,q) = F(p,q+a)
\end{equation}
provided that the result, i.e., $F(p,q+a)$, is also a function on  $\mathbbm{R} \times {\cal Z}$ .  For simplicity we assume that $a>0$. When $0<a<1$, $S_a$ is defined provided that $F(p,q)$ vanishes  for $q$ not satisfying $3+a\leq q \leq 4$ .  As a second example, if $F(p,q)$ is nonzero for only $q=3$, then $S_a$ is defined for $a$ $=$ $1$, $2$, and $3$.  There are other nonvanishing examples, but these two examples serve to illustrate the main idea.  And that main idea is:  For suitable functions on the constrained classical phase space the translation transformation $S_a$ is defined for selected values of the translation variable consistent with the structure of $\mathbbm{R} \times {\cal Z}$.  Of course, no attempt to define $S_a$ in terms of a generator and power series
[as in (\ref{qwer})] is possible in the general case.  

Now let us turn our attention to the physical Hilbert space.  We approach this discussion in two steps:  first, at the level of ${\cal K}_n$, and second at the level of ${\tilde {\cal K}}_n$.  As observed in Sec. 3, physical observables ${\cal O}$ must obey the identity 
\begin{equation}
[\mathbbm{E}, {\cal O}] = 0.
\end{equation}
 In particular, $[\mathbbm{E},Q^E]=0$, where  
\begin{equation}
Q^E=\mathbbm{E}Q\mathbbm{E},
\end{equation}
but since $\mathbbm{E}=\mathbbm{E}(-\delta<R(Q)<\delta)$ it follows that 
\begin{equation}
Q^E=\mathbbm{E}^2Q=\mathbbm{E}Q = Q\mathbbm{E}
\end{equation}
as well.  Even simpler is the case of the identity  for which,
\begin{equation}
\mathbbm{1}^E=\mathbbm{E}.
\end{equation}
With regard to the preclassical (still with $\hbar >0$) phase space function associated with the operator $\mathbbm{1}^E$ , we are led to the equation 
\begin{equation}
\frac{\<p,q|\mathbbm{E}\mathbbm{1}^E\mathbbm{E}|p,q\>}{\<p,q|\mathbbm{E}|p,q\>}=1,
\end{equation}
since the classical limit of the unit operator on any Hilbert space should be unity.  This simple remark is made to emphasize the fact that such expectation values will involve both a numerator and a denominator as illustrated above.  

We now consider the corresponding quotient for $Q^E$, namely
\begin{equation}
\frac{\<p,q|\mathbbm{E}Q\mathbbm{E}|p,q\>}{\<p,q|\mathbbm{E}|p,q\>},
\end{equation}
an intermediate expression which serves principally as motivation.  We shall not evaluate this expression explicitly but pass immediately to the quotient composed of terms made separately from ${\tilde {\cal K}_n}$ kernels in which the limit $\delta \rightarrow 0$ has already been taken.  Hence, we are led to the equation
\begin{equation}
q^P(p,q;\hbar) \equiv \frac{\Sigma q_lA_le^{-(q_l-q)^2/\hbar}+B\int^4_3 x e^{-(x-q)^2/\hbar}dx}{\Sigma A_le^{-(q_l-q)^2/\hbar}+B\int^4_3  e^{-(x-q)^2/\hbar}dx}
\end{equation}
where the superscript $P$ refers to ``physical". Here we have reinserted $\hbar$ into our expressions and at the same time generalized them by allowing arbitrary positive constants $A_l$ and $B$ permitting alternative (finite) rescaling of each ${\tilde {\cal K}_n}$ entering ${\tilde {\cal K}}$.  Such rescalings have the effect of generating similarity transformations among such Hilbert spaces.  For our standard example $q_1 =0$, $q_2=1$, $q_3=2$, $q_4=3$, and $q_5=4$ .  As it stands, $q^P(p,q;\hbar)$ depends on $\hbar$; let us determine the classical limit that emerges as $\hbar \rightarrow 0$.  A moments reflection will show that the classical limit of the physical (observable) position variable is given by 
\begin{equation}
q^P(p,q) \equiv \lim_{\hbar\to 0} q^P(p,q;\hbar),
\end{equation}
where
\begin{equation}
q^P(p,q) = 
\begin{cases}
q_1 (=0),& q<1/2,\\
q_2 (=1),& 1/2<q<3/2,\\
q_3 (=2),& 3/2<q< 5/2, \\
q_4 (=3),& 5/2<q\leq 3,\\
q  ,& 3 <q < 4,\\
q_5 (=4), & 4 \leq q.
\end{cases}
\end{equation}
These values hold independently of whatever choice is made for $A_l$ and $B$!  It is noteworthy that the set of allowed values for $q^P(p,q)$ exactly coincides with the zero set ${\cal Z} = \{q=0,q=1,q=2,3\leq q \leq 4\}$  for our model problem.

Observe that each discrete value  ($q_1,q_2, q_3,q_4, q_5$) occurs for a set of nonzero measure in the original phase space.  The excluded points in $q$, such as $q=1/2$, lie exactly midway between two discrete values.  Consequently
\begin{equation}
q^P(p,1/2) \equiv \frac{q_1A_1+q_2A_2}{A_1+A_2}=\frac{A_2}{A_1+A_2}\;,
\end{equation}
a weighted average of the two equally-distant, nearest discrete levels.  Such anomalous values are inevitable when distinct, discrete coordinate values are allowed; note that no such behavior arises at $q=7/2$.  Since such values which lie outside the set ${\cal Z}$ occur on a set of measure zero in the original phase space, they may safely be ignored. 

Equation (\ref{mo1}) shows that the classical limit of the physical momentum $p^P(p,q)=p$.  More interesting is the quantum analog of the translation operator and its classical limit.  In the unconstrained Hilbert space the unitary operator 
\begin{equation}
U_a = e^{iaP/\hbar}
\end{equation}
acts to actively translate the operator $Q$ in the manner
\begin{equation}
U_aQU_a^\dagger = Q+a.
\end{equation}
The observable parts of $U_a$ and $U_a^\dagger$ are given by 
\b
U_a^E &=&\mathbbm{E}U_a\mathbbm{E}, \\
{U_a^\dagger}^E &=&\mathbbm{E}U_a^\dagger\mathbbm{E}
\e
and the Hermitian operator
\begin{equation}
W_a\equiv U_a^E{U_a^\dagger}^E = \mathbbm{E}U_a\mathbbm{E}U_a^\dagger\mathbbm{E}\end{equation}
denotes the remnant of the translation symmetry in the regulated version of the physical Hilbert space.  Observe that 
\begin{equation}
W_a = \mathbbm{E}(-\delta<R(Q)<\delta)\; \mathbbm{E}(-\delta<R(Q+a)<\delta).
\end{equation}
For very small $\delta>0$ it is clear that this projection operator is essentially seeking common support in the two sets  ${\cal Z} = \{q=0,q=1,q=2,3\leq q \leq 4\}$ and   ${\cal Z}_a = \{q=-a,q=1-a,q=2-a,3-a\leq q \leq 4-a\}$ .  For example, if $0<a<1$, then for our standard example
\begin{equation}
 \mathbbm{E}U_a\mathbbm{E}U_a^\dagger\mathbbm{E}= \mathbbm{E}(3+a \leq Q \leq 4).
\end{equation}
If $a=1$ then $W_1$ is nonzero near $Q= 0,1,2,$ and $3$ with some small spreading determined by $\delta$.

Passing from the level of ${\cal K}_n$ to the level of ${\tilde {\cal K}_n}$ we are led to the equation
\begin{equation}
w^P_1(p,q;\hbar) \equiv \frac{\Sigma'_l A_le^{-(q_l-q)^2/\hbar}}{\Sigma_l A_le^{-(q_l-q)^2/\hbar}+B\int^4_3  e^{-(x-q)^2/\hbar}dx}
\end{equation} 
for the preclassical $(\hbar >0)$ phase space symbol associated with the operator $W_1$.  The prime on the numerator sum signifies that the term for $q_5 =4$ is omitted from the sum.  The classical limit of this expression leads to 
\begin{equation}
w_1^P(p,q) \equiv \lim_{\hbar\to 0} w_1^P(p,q;\hbar), 
\end{equation}
where
\begin{equation}
w_1^P(p,q)= 
\begin{cases}
1,& q<1/2,\\
1,& 1/2<q<3/2,\\
1,& 3/2<q\leq 3, \\
0, & 3<q.
\end{cases}
\end{equation}
This evaluation is interpreted as a validation of the nonvanishing translation transformation $S_1$ that applies in the constrained classical phase space and which arises from a corresponding translation transformation in the quantum mechanical physical Hilbert space.  A similar calculation for an inappropriate translation value, such as $a=3/2$, would lead to a 
vanishing classical expression.  

In summary, the successful recovery of allowed translation transformation values for the classical reduced phase space from the quantum theory adds further credibility to our formulation.  We emphasize that by applying the formalism of the projection operator approach to highly irregular quantum constraints we have found no evidence of a requirement for superselection sectors proposed earlier  nor a restricted form of the classical limit of the reduced quantum theory using a qualitatively different formulation to deal with the quantum constraints \cite{lou}, both aspects of which would show up as well in our simplified model since they only involve the study of a constraint that is both regular and irregular.

\section{Acknowledgments}
Jan Govaerts and Jorma Louko are thanked for their helpful comments regarding the paper and irregular constraints in dynamical systems.

\end{document}